\documentclass[twocolumn,showpacs,amsmath,amssymb,prb,aps]{revtex4}

\usepackage{graphicx}
\usepackage{dcolumn}
\usepackage{bm}
\usepackage{amsmath}
\usepackage{amsfonts}
\usepackage{amssymb}
\usepackage{MnSymbol}

\begin{document}

\preprint{APL/NbN resonators}

\title{In-situ measurement of the permittivity of helium using microwave NbN resonators}

\author{G. J. Grabovskij}
\author{L. J. Swenson}
\author{O. Buisson}
\author{C. Hoffmann}
\author{A. Monfardini\footnote{Electronic mail : alessandro.monfardini@grenoble.cnrs.fr}}

\affiliation{Institut N\'{e}el, CNRS \& Universit\'{e} Joseph Fourier, 25 rue des Martyrs, 38042 Grenoble Cedex 9, France}
%exclude Centre National de la Research Scientifique to save room

\author{J.-C. Vill\'{e}gier}

\affiliation{CEA-INAC, CEA-Grenoble, 17 rue des Martyrs, 38054 Grenoble Cedex 9, France}

\date{\today}

\begin{abstract}
    By measuring the electrical transport properties of superconducting NbN quarter-wave resonators in direct contact with a helium bath, we have demonstrated a high-speed and spatially sensitive sensor for the permittivity of helium.  In our implementation a $\sim$10$^{-3}$ mm$^3$ sensing volume is measured with a bandwidth of 300 kHz in the temperature range 1.8 to 8.8 K.  The minimum detectable change of the permittivity of helium is calculated to be $\sim$6$\times$$10^{-11}$ $\epsilon_0$/Hz$^{1/2}$ with a sensitivity of order $10^{-13}$ $\epsilon_0$/Hz$^{1/2}$ easily achievable.  Potential applications include operation as a fast, localized helium thermometer and as a transducer in superfluid hydrodynamic experiments.
\end{abstract}


\maketitle

With their ease of fabrication, high quality factors and the wide availability of measurement electronics, superconducting microwave resonators have recently emerged as ideal systems for studying mesoscopic physics.  Their applications are widespread and include electromagnetic radiation detectors, \cite{day:817, doyle:530} electromechanical transducers, \cite{flowers-jacobs:096804, regal:555} and quantum computing circuit elements.\cite{wallraff:162, vion:886, hofheinz:310}  In realizing a superconducting resonator, it is often convenient to use a high-frequency transmission line with a planar geometry, enabling straight forward integration with other on-chip devices.  Typical designs include microstrip or coplanar waveguide (CPW).  However, these geometries are very sensitive to their environment and it is necessary to minimize dielectric loss in order to achieve optimal performance.  This can be accomplished, for example, by utilizing a low loss dielectric such as sapphire for the support substrate and performing the experiment in vacuum.

While dissipation of energy in the surrounding dielectric usually limits performance of a superconducting resonator, it is also possible to utilize this feature to probe intrinsic material properties.  For example, by fabricating a series of half-wavelength CPW resonators coated with a variety of dielectrics, the high-frequency, low-temperature and low-energy properties of common insulating materials have recently been reported. \cite{o'connell:112903}  In a similar study, quarter-wavelength CPW resonators coated with varying thicknesses of SiO$_{x}$ were used to investigate the noise properties of thin film dielectrics. \cite{barends:223502}

\begin{figure}
\includegraphics[width=7.8cm]{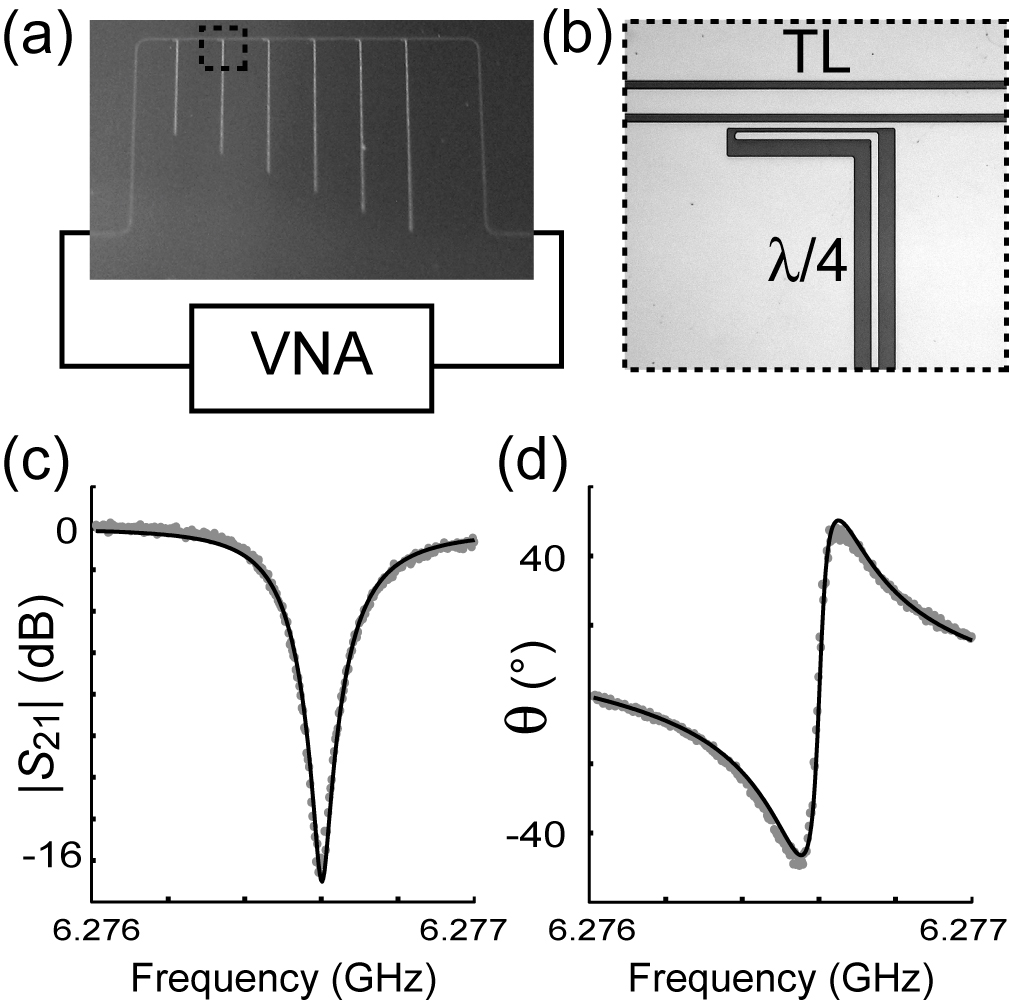}
 \caption{Measurement circuit. (a) A vector network analyzer (VNA) with an output power of -50 dBm was used to measure the transmission coefficient $S_{21}$.  The six resonators shown range in length from 2240 to 4430 $\mu$m with frequencies $\sim$12 to 6.3 GHz.  (b) The center conductor width is 19 $\mu$m for the main transmission line (TL) and 5 $\mu$m for the quarter-wavelength resonator.  The channel width is 33 $\mu$m in both cases. The end of the resonator (not shown) is shorted to ground.  The coupling capacitance is estimated to be 4 fF. (c) Typical resonance curve $|S_{21}|$ measured at 2.2 K.  The off-resonance amplitude is adjusted to 0 dB to account for attenuation in the measurement system.  The black line is a fit which takes into account the resonance frequency, the internal and external quality factors, and parasitic impedances introduced by the measurement circuitry. (d)  Phase $\theta$ measured simultaneously with the amplitude data in (c).}
\label{fig:figure1}
\end{figure}

\begin{figure}[b]
\includegraphics[width=7.8cm]{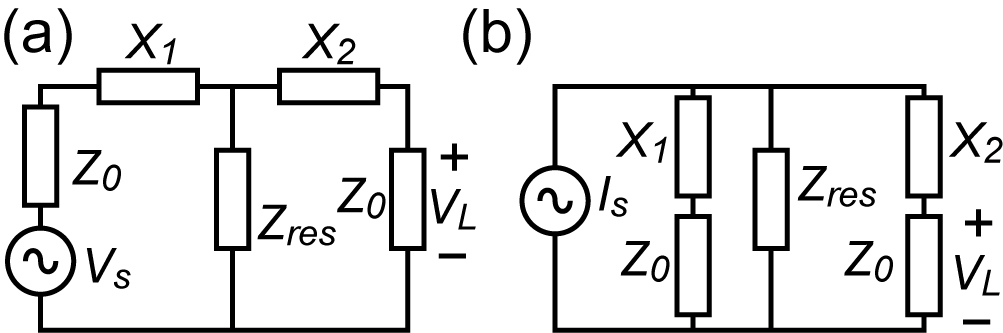}
 \caption{(a) Th\'{e}venin and (b) Norton equivalent circuits.}
\label{fig:figure2}
\end{figure}

Here we demonstrate the ability of CPW superconducting resonators to sensitively detect changes in the permittivity of surrounding helium at low temperatures (1.8 to 8.8 K).  Previous measurements have relied on large-volume cavity resonances ($\sim$1 mm$^3$) or audio-frequency measurements of parallel-plate capacitors ($f < 10$ kHz), thus excluding high-speed localized measurements.\cite{greenkemper:89, niemela:1}  In contrast, the small sensing volume ($\sim$10$^{-3}$ mm$^3$), open geometry and tunable bandwidth of CPW resonators are ideally suited to fast, localized measurements.  The high spatial sensitivity can also be made to extend over large detection volumes by frequency-multiplexing an array of resonators.  This could enable, for example, measurement of local fluctuations in quantum hydrodynamic experiments similar to previous measurements, \cite{roche:66002} but with increased volumetric and temporal resolution.

A diagram of our experiment is shown in Fig.\ \ref{fig:figure1}.  The fabrication of quarter-wave superconducting resonators has been previously reported.\cite{mazin:2004, barends:257002} Briefly, 150 nm of B1-cubic NbN phase was sputter-deposited epitaxially on a 430-$\mu$m-thick M-plane sapphire substrate.\cite{espiau:232501}  This was then patterned using photo-lithography and subsequent reactive-ion etching in a SF$_6$/O$_2$ plasma.  NbN was chosen for its high $T_c$ of $\sim$15 K to enable a large temperature range for measurement.  The wafer was then mounted on a copper-clad board and wire-bonded to a 50-$\Omega$ transmission line.  Multiple bonds were made in order to reduce the parasitic reactance of the wire bonds.  The board was enclosed in a custom-designed microwave package and attached to stainless-steel coaxial cables at the cold stage of a helium cryostat.  No vacuum space was used in order to allow direct contact between the helium bath and the NbN resonators.

The cryostat was immersed in liquid helium and electrical transport data was taken with a vector network analyzer.  The amplitude $|S_{21}|$ for a typical transmission measurement is shown in Fig.\ \ref{fig:figure1}(c) and the corresponding phase $\theta$ in Fig.\ \ref{fig:figure1}(d).  Inspection of the measured data reveals an asymmetry in the transmission $S_{21}$ about the resonant frequency $f_0$. The usual explanation for this asymmetry is a parasitic inductance associated with the bond wires. \cite{doyle:530, krems:178}  To account for this effect we have modeled our resonant circuit as a shunt impedance $Z_{res}$ connected to the source and detection ports of the network analyzer ($Z_{0}$ = 50 $\Omega$) by parasitic reactances $X_1$ and $X_2$.  $Z_{res}$ includes the coupling capacitance between the resonator and transmission line in addition to the intrinsic resonator impedance.  The Th\'{e}venin and Norton equivalent circuits are shown in Fig.\ \ref{fig:figure2}.  Using this circuit model, we can calculate the complex scattering parameter
\begin{equation}\label{eq01}
    S_{21} = \frac{2Z_{res}Z_0}{Z_{res}[2Z_0 + j(X_1 + X_2)] + (Z_0 + jX_1)(Z_0+jX_2)},
\end{equation}
where $X_1$ and $X_2$ are assumed to be frequency independent for a given resonator as the fractional change in the frequency ($\delta f$/$f_0$) is small over the measurement bandwidth.  The loaded resonator impedance at a frequency $f$ is given by\cite{pozar:2005, mazin:2004}
\begin{equation}\label{eq02}
    Z_{res} = \frac{Z_0 Q_e}{2 Q_i} \left[1 + 2 j Q_i \frac{(f-f_0)}{f_0}\right],
\end{equation}
where $Q_i$ is the intrinsic quality factor of the resonator and $Q_e$ is the external quality factor due to coupling with the measurement electronics.  The measurement's loaded quality factor $Q_L$ is given by $1/Q_L$ = $1/Q_i$ + $1/Q_e$.

We have used Eq.\ (\ref{eq01}) to fit the data for our resonators.  While the parasitic reactances do not change significantly in the bandwidth of a single resonator, they will in general depend on frequency and hence can be different for each resonator.  However, for any given resonator these fit parameters are found to not vary substantially with temperature and in nearly all cases, $X_1=X_2=X$.  This matches well with the assumption that the asymmetry in $S_{21}$ about $f_0$ is largely due to the bond wires. For the resonance shown in Fig.\ \ref{fig:figure1} a typical value of $X=3$ $\Omega$ was found.

\begin{figure}
\includegraphics[width=7.8cm]{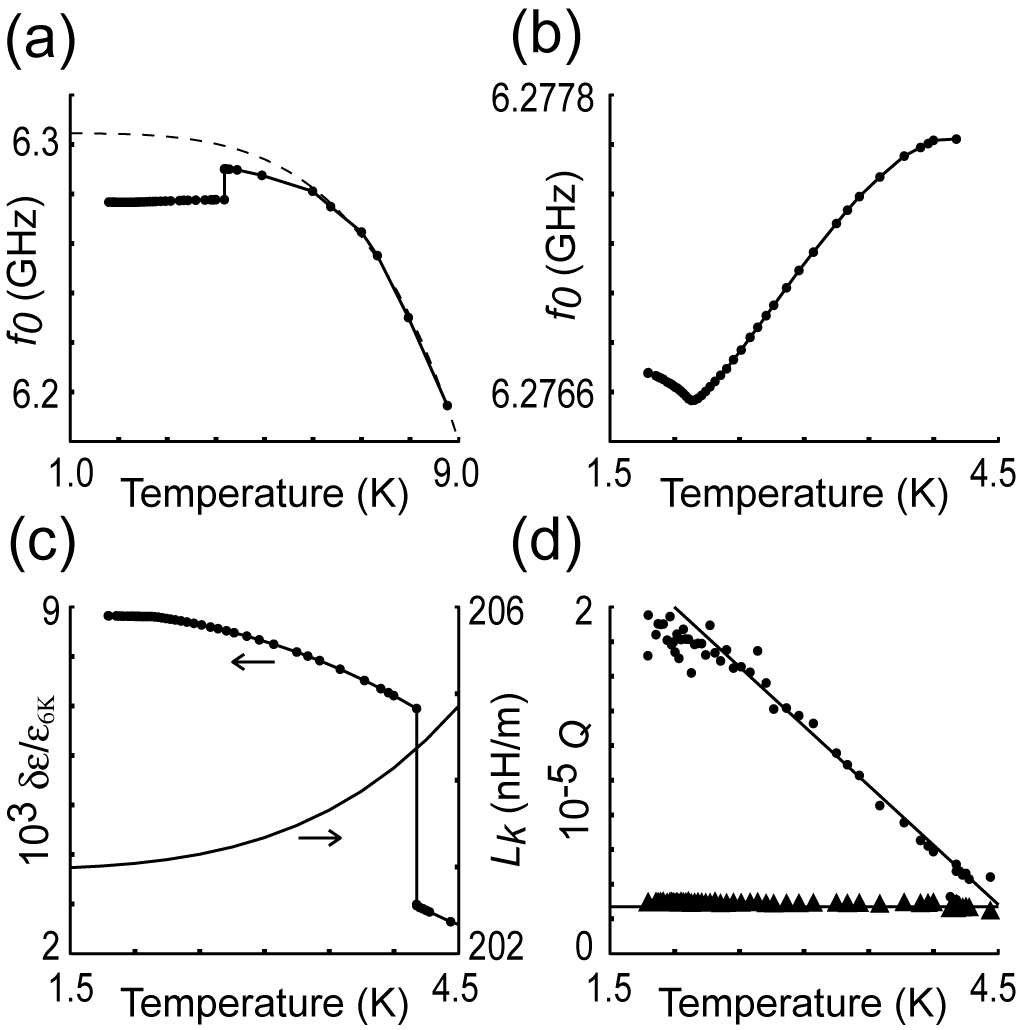}
 \caption{(a)  Measured shift in the resonant frequency $f_0$ with temperature.  The dashed line is a fit of the high-temperature data ($T > 6$ K) where the temperature-dependent kinetic inductance $L_k$ dominates the response.  Discrepancies between the fit and the measured data at low temperature are due to changes in the dielectric permittivity.  (b)  Measured shift in $f_0$ below  4.2 K.  The superfluid transition is prominent at 2.2 K. (c)  Extracted kinetic inductance $L_K$ and fractional change in permittivity $\delta \epsilon/\epsilon_{6K}$.  (d) Internal ($Q_i$, $\bullet$) and external ($Q_e$, $\filledmedtriangleup$) quality factors.  The solid lines are guides to the eye.  In this case, the external quality factor limits measurement speeds to $\sim$ 3 $\mu$s.}
\label{fig:figure3}
\end{figure}

The extracted resonance frequency $f_0$ is shown as a function of temperature for a resonator of length 4540 $\mu$m in Fig.\ \ref{fig:figure3}(a) and below 4.2 K in Fig.\ \ref{fig:figure3}(b).  The structure of this data is governed by\cite{pozar:2005, barends:257002}
 \begin{equation}\label{eq03}
    f_0 \cong \frac{1}{4 l \, \sqrt{(L_K+L_G) C(\epsilon)}},
 \end{equation}
where $l$ is the resonator's length, $\epsilon$ is the effective permittivity of the resonator, and $L_K$, $L_G$ and $C$ are the resonator's kinetic inductance, geometric inductance and capacitance respectively, each expressed as a reactance per unit length.  The average kinetic inductance per unit length is given by \cite{schmidt:1997}
\begin{equation}\label{eq04}
    L_K = \frac{\mu_0 \lambda(T)^2}{l} \int \frac{j_s^2}{I^2} dV,
\end{equation}
where $j_s$ is the local supercurrent density; $I$ is the total current; $\lambda(T)$, the London penetration depth, is given by the empirical formula
$\lambda(T) = \lambda(0)/[1-(T/T_c)^4]^{1/2}$;
%\begin{equation}\label{eq05}
%    \lambda(T) = \frac{\lambda(0)}{[1-(T/T_c)^4]^{1/2}};
%\end{equation}
and the integral is taken over the volume of the superconductor.  Since the magnetic penetration depth of NbN is of the order of our film thickness ($\lambda(0) \sim 350$ nm),\cite{lamura:104507} it is reasonable to assume a uniform supercurrent density if we neglect fringing effects.  In this case, the kinetic inductance reduces to $L_k =  \mu_0 \lambda(T)^2/A$, where A is the cross-sectional area of the NbN film.

At temperatures appreciably above the boiling point of helium and at standard pressure, the dielectric permittivity is temperature insensitive.  In this regime, the measured change in $f_0$ is dominated by $L_K$.  A fit of Eq.\ (\ref{eq03}) to the data above 6 K  assuming constant permittivity is shown in Fig.\ \ref{fig:figure3}(a). The temperature-independent geometric inductance $L_G$ and capacitance $C$ were determined utilizing finite-element software to be 661 nH/m and 93 pF/m, respectively.  From this fit, we find $\lambda(0) = 348$ nm and $T_c = 14.5$ K for the NbN film, consistent with previous measurement. \cite{lamura:104507}  For temperatures below 6 K, the data deviates from the fit due to changes in the dielectric.  The ratio of the fitted curve ($f_0'$) to the measured data ($f_0$) allows determination of the permittivity through $\delta \epsilon/\epsilon_{6K} = (f_0'/f_0)^2 - 1$, where $\delta \epsilon/\epsilon_{6K}$ is the fractional change in the permittivity measured from the permittivity at 6 K.  Fig.\ \ref{fig:figure3}(c) displays the temperature dependence of $L_k$ and $\delta \epsilon/\epsilon_{6K}$ determined from this analysis.

It is useful to estimate the sensitivity of our NbN quarter-wave resonators to the permittivity of helium $\epsilon_{He}$.  From Eq.\ (\ref{eq03}), at low temperatures $f_0 \propto 1/\sqrt{\epsilon}$ and thus for small changes in the permittivity, $\delta f_0/f_0 = -\delta \epsilon/2\epsilon$.  For CPW, the effective permittivity is approximately the average of the environmental and substrate permittivities, or in our case, $\epsilon \sim (\epsilon_{He} + \epsilon_{Al_2O_3})/2$.  In terms of small changes in the permittivity of helium, $\delta f/f_0 \sim -\delta \epsilon_{He}/4\epsilon$.  The change in the measurement phase for small changes in frequency about resonance is given by the relationship $\delta \theta = [2 Q_i^2/(Q_i + Q_e)](\delta f/f_0)$.  The phase uncertainty of an amplifier with a noise temperature $T_n$ is $\sigma_{\theta} = \sqrt{(k_B T_n)/2 P_s}$ (radians/Hz$^{1/2}$), where $P_s$ is the measurement signal power.\cite{mazin:2004}  Combining these, we find for the sensitivity
\begin{equation}\label{eq06}
    \delta \epsilon_{He}^{min} \cong 2 \epsilon \sqrt{\frac{k_B T_n}{2 P_s}}\left(\frac{Q_i+Q_e}{Q_i^2}\right),
\end{equation}
assuming the limiting noise contribution is from the first stage amplifier.  Using $Q_i = 10^5$, $Q_e = 2\times10^4$, $P_s = -50$ dBm, $\epsilon = 5.5 \epsilon_0$, and an amplifier noise temperature of 290 K, we calculate a sensitivity of $\sim$$6\times10^{-11}$ $\epsilon_0$/Hz$^{1/2}$ for our measurement.  By utilizing a cryogenic amplifier, increasing the measurement signal power, and decreasing the coupling, we estimate a sensitivity of order $10^{-13}$ $\epsilon_0$/Hz$^{1/2}$ can easily be achieved.

In conclusion, we have demonstrated a small-volume ($\sim$10$^{-3}$ mm$^3$), high-speed ($\tau \sim$ 3 $\mu$s) technique for in-situ measurement of the permittivity of helium.  Potential applications of this device include operation as an exquisite helium thermometer or as a spatially-sensitive transducer in superfluid hydrodynamic experiments.

This work was supported by the EuroSQIP project and RTRA ``Nanosciences Grenoble.''

\end{document}